\documentclass[pra, 12 pt]{revtex4}

\begin{document}

\title{Modified Newtonian dynamics from acceleration fluctuations}

\author{Thomas F. Jordan}
\affiliation{Physics Department, University of Minnesota, Duluth, Minnesota 55812}
\email[email: ]{tjordan@d.umn.edu}

\begin{abstract}
A speculative mathematical model is used to generate the modified Newtonian dynamics called MOND from fluctuations of the number of quanta of quantized acceleration.  The one new parameter can be chosen either to make the transition to modification comparable to that obtained from the functions used to fit data with MOND, or to make the modification at larger accelerations comparable in magnitude to the unexplained accelerations of Pioneer 10 and 11.
\end{abstract}
\maketitle

The idea of the modified Newtonian dynamics called MOND is that accelerations are changed when they drop below
\begin{equation}
  \label{eq:a0}
  a_0=1.2 \times 10^{-8} cm/s^2.
\end{equation}
If the unmodified acceleration is $g$ and the modified acceleration is $a$, then
\begin{equation}
  \label{eq:a1}
  a=g \quad {\mbox{for}} \quad g >> a_0 
\end{equation}
and
\begin{equation}
  \label{eq:a2}
  a= \sqrt{a_0} \, \sqrt{g} \quad {\mbox{for}} \quad g << a_0.
\end{equation}
It is assumed that the transition is made with a smooth function $\mu$ that gives
\begin{equation}
  \label{eq:a3}
  a \, \mu (a/a_0)=g.
\end{equation}
This function must have the properties that $\mu(x)$ is $1$ when $x$ is large
compared to $1$ and $\mu(x)$ is $x$ when $x$ is small compared to $1$, but its form is not specified.  The forms most often used in calculations are
\begin{equation}
  \label{eq:a5}
  \mu(x) = \frac{x}{\sqrt{1+x^2}}
\end{equation}
and
\begin{equation}
  \label{eq:a6}
  \mu(x) = 1-e^{-x}.
\end{equation}
This program gives remarkable fits to galaxy rotation curves and other astrophysical data.\cite{sanders}

In speculating about what could cause this, we can consider a mathematical model without understanding what the physical basis for the model might be. In doing this, we run the risk of not being able to tell whether a model is as significant as one of Kepler's laws or as silly as Kepler's fantasy of the five perfect solids separating the orbits of six planets.

Suppose the acceleration is quantized in units of $a_0$.  The unmodified acceleration is 
\begin{equation}
  \label{eq:a7}
  g = a_0 \bar{n}
\end{equation}
where $n$ is the number of quanta in a part of the acceleration process and $\bar{n}$ is the average of $n$.  When the acceleration is large compared to $a_0$, the average of $n$ is large and is all that matters.  Then $g$ is the acceleration.  We are accustomed to that. However, suppose the number $n$ fluctuates so that
\begin{equation}
  \label{eq:a8}
  \bar{n^2} = \bar{n}^2 + \bar{n}
\end{equation}
as in a Poisson distribution, and suppose the acceleration depends on the square of $n$, so the actual acceleration is
\begin{equation}
  \label{eq:a9}
  a=a_0(\bar{n^2})^{\frac{1}{2}}= a_0 (\bar{n}^2 + \bar{n})^{\frac{1}{2}}.
\end{equation}
Then
\begin{equation}
  \label{eq:a10}
  a=a_0\left[ \left( \frac{g}{a_0} \right)^2 + \frac{g}{a_0} \right]^{\frac{1}{2}} = g \left(1+\frac{a_0}{g} \right)^{\frac{1}{2}} = \sqrt{a_0}\sqrt{g} \left(1 + \frac{g}{a_0} \right)^{\frac{1}{2}}.  
\end{equation}
This clearly satisfies equations (\ref{eq:a1}) and (\ref{eq:a2}). The transition, the modification, occurs because the fluctuations of $n$ enter and become dominant as $g$ drops below $a_0$.

The same can be said if
\begin{equation}
  \label{eq:a11}
  g=a_0 \bar{n}^{2p}
\end{equation}
and
\begin{equation}
  \label{eq:a12}
  a=a_0(\bar{n^2})^{p} = a_0(\bar{n}^2+\bar{n})^{p}
\end{equation}
so
\begin{equation}
  \label{a13}
  a=a_0\left[ \left( \frac{g}{a_0} \right)^{\frac{1}{p}} + \left(\frac{g}{a_0} \right)^{\frac{1}{2 p}} \right]^{p} = g \left[1+ \left(\frac{a_0}{g} \right)^{\frac{1}{2p}} \right]^{p} = \sqrt{a_0}\sqrt{g} \left[1 + \left(\frac{g}{a_0}\right)^{\frac{1}{2p}} \right]^{p}
\end{equation}
with $p$ a positive power.  We will find that results are significantly changed when $p$ is only a little different from $1/2$.

For a range of values of $p$, the rate of convergence to $\sqrt{a_0} \sqrt{g}$ for small $g$ is comparable to that obtained from the functions (\ref{eq:a5}) and (\ref{eq:a6}) for $\mu$ that are used to fit data.  Suppose $g$ is small compared to $a_0$.  Then so is $a$.  Using equation (\ref{eq:a6}) for $\mu$ in equation (\ref{eq:a3}) gives
\begin{eqnarray}
  \label{eq:a14}
  a \left[ \frac{a}{a_0} - \frac{1}{2} \left( \frac{a}{a_0} \right)^2 \right] & \simeq & g,  \nonumber \\
a^2 \left( 1 - \frac{1}{2} \frac{a}{a_0} \right) & \simeq & a_0 g, \nonumber \\
 a \simeq \sqrt{a_0} \sqrt{g} \left( 1 + \frac{1}{2} \frac{a}{a_0} \right)^{\frac{1}{2}}& \simeq &\sqrt{a_0} \sqrt{g} \left( 1 + \frac{1}{4} \frac{a}{a_0} \right)  \nonumber, \\
\frac{a-\sqrt{a_0} \sqrt{g}}{\sqrt{a_0} \sqrt{g}} &\simeq& \frac{1}{4} \frac{a}{a_0}\simeq \frac{1}{4} \left( \frac{g}{a_0} \right)^{\frac{1}{2}}.
\end{eqnarray}
Using equation (\ref{eq:a5}) for $\mu$ in equation (\ref{eq:a3}) gives
\begin{eqnarray}
  \label{eq:a15}
  a \left[ \frac{a}{a_0} - \frac{1}{2} \left( \frac{a}{a_0} \right)^3 \right] & \simeq & g,  \nonumber \\
\frac{a-\sqrt{a_0} \sqrt{g}}{\sqrt{a_0} \sqrt{g}} &\simeq& \frac{1}{4} \left(\frac{a}{a_0} \right)^2 \simeq \frac{1}{4}  \frac{g}{a_0}.
\end{eqnarray}
In comparison, equation (\ref{a13}) gives
\begin{equation}
  \label{eq:a16}
  \frac{a-\sqrt{a_0} \sqrt{g}}{\sqrt{a_0} \sqrt{g}} \simeq p \left( \frac{g}{a_0} \right)^{\frac{1}{2 p}}
\end{equation}
which for
\[
  \begin{array}{lllllll}p & = & \quad\frac{1}{4},&  \quad\frac{1}{3}, &  \;\frac{1}{2},&  \quad\frac{2}{3},&  \quad\frac{3}{4}, \\
{\mbox is} & &  \frac{1}{4}\left( \frac{g}{a_0} \right)^{2}, \qquad & \frac{1}{3}\left( \frac{g}{a_0} \right)^{\frac{3}{2}}, \qquad &  \frac{1}{2}\frac{g}{a_0}, \qquad &\frac{2}{3}\left( \frac{g}{a_0} \right)^{\frac{3}{4}}, \qquad &\frac{3}{4}\left( \frac{g}{a_0} \right)^{\frac{2}{3}}.
\end{array}
\]

On the other hand, convergence of the acceleration to $g$ for large $g$ can be slower than that obtained from the functions (\ref{eq:a5}) and (\ref{eq:a6}) for $\mu$.  The difference between $a$ and $g$ can stay large enough to be tested by observations in the solar system.  The unexplained acceleration of the Pioneer 10 and 11 spacecraft toward the sun, recently estimated very carefully,\cite{anderson} would give
\begin{equation}
  \label{eq:a17}
  a-g=(8.74 \pm 1.33) \times 10^{-8} cm/s^2.
\end{equation}
This is roughly constant over distances between 14 and 44 AU from the sun. It is also estimated \cite{anderson} that an acceleration larger than $0.1 \times 10^{-8} \; cm/s^2$ could not have gone undetected in observations of the orbits of the Earth and Mars.  Our equation (\ref{eq:a10}) gives
\begin{equation}
  \label{eq:a18}
  a-g \simeq \frac{1}{2} a_0
\end{equation}
for $g$ large compared to $a_0$.  Using the value (\ref{eq:a0}) for $a_0$, we see this is $7$ times too small to be what is seen for Pioneer 10 and 11 and $6$ times too large to be unseen for the Earth and Mars.

This is where a small difference in $p$ can change the results.  For $g$ large compared to $a_0$, equation (\ref{a13}) gives
\begin{equation}
  \label{eq:a19}
  a-g \simeq p a_0^{\frac{1}{2p}}g^{1-\frac{1}{2 p}}= a_0 p \left( \frac{a_0}{g} \right)^{\frac{1}{2p}-1}.
\end{equation}
For gravity from the Sun on the orbit of Mars, $g$ is about $0.25 \; cm/s^2$ and $a_0/g$ is about $4.7 \times 10^{-8}$, so if $p$ is $4/9$, then
\begin{equation}
  \label{eq:a20}
  a-g \simeq a_0 \frac{4}{9}(4.7 \times 10^{-8})^{\frac{1}{8}} \simeq \frac{1}{19}a_0
\end{equation}
which may be too small to have been observed.

For Pioneer 10 and 11 at distances around 30 AU from the Sun, $g$ is about $6 \times 10^{-4} cm/s^2$ and $a_0/g$ is about $2 \times 10^{-5}$, so if $p$ is $5/8$ then
\begin{equation}
  \label{eq:a21}
  a-g \simeq a_0 \frac{5}{8} ( 2 \times 10^{-5})^{-\frac{1}{5}} \simeq 7 a_0
\end{equation}
which is close to what is observed.  However, when the distance of the spacecraft from the Sun increases by a factor of $3$, so $g$ decreases by a factor of $9$,  if $p$ is $5/8$ then $a-g$ should decrease by a factor of $9^{1/5}$, which is about $1.55$. The data shows perhaps a bit of change in that direction, but not that much.\cite{anderson}

Smaller differences between $a$ and $g$ are obtained from the functions (\ref{eq:a5}) and (\ref{eq:a6}) for $\mu$.  For large $x$, equation (\ref{eq:a5}) gives
\begin{equation}
  \label{eq:a22}
  1 - \mu(x) = \frac{\sqrt{1+x^2}-x}{\sqrt{1 +x^2}}=  \frac{\sqrt{1/x^2+1}-1}{\sqrt{1/x^2+1}} \simeq \frac{1}{2 x^2}
\end{equation}
so for $g$ large compared to $a_0$, using the function (\ref{eq:a5}) for $\mu$ in equation (\ref{eq:a3}) gives
\begin{equation}
  \label{eq:a23}
  a-g = (1-\mu) a \simeq \frac{1}{2}\left( \frac{a_0}{a} \right)^2 a \simeq \frac{1}{2} \frac{a_0}{g} a_0.
\end{equation}
Using the function (\ref{eq:a6}) for $\mu$ in equation (\ref{eq:a3}) gives
\begin{equation}
  \label{eq:a24}
   a-g = (1-\mu) a = e^{-a/a_0}a \simeq e^{-g/a_0} \frac{g}{a_0}a_0
\end{equation}
for $g$ large compared to $a_0$.  These are too small to be observed in the solar system.  Even for Pioneer 10 and 11 out at 70 AU from the Sun, $g$ is about $(0.59 \; cm/s^2)/(70)^2$ and $a_0/g$ is about $10^{-4}$, so equations (\ref{eq:a23}) and (\ref{eq:a24}) would make the difference between $a$ and $g$ orders of magnitude smaller than what is observed.

When $p$ is $1/4$, equation (\ref{eq:a19}) gives
\begin{equation}
  \label{eq:a25}
  a-g \simeq \frac{1}{4}\frac{a_0}{g} a_0.
\end{equation}
This is comparable to equation (\ref{eq:a23}).  Convergence of the acceleration to $g$ for large $g$ when $p$ is $1/4$ is comparable with that obtained from the function (\ref{eq:a5}) for $\mu$.  We have seen that for a range of values of $p$, convergence of the acceleration to $\sqrt{a_0} \, \sqrt{g}$ for small $g$ is comparable to that obtained from the functions (\ref{eq:a5}) and (\ref{eq:a6}) for $\mu$. For a suitable value of $p$, perhaps $1/4$, our model might be expected to fit the data that MOND can fit.  To see it as possibly extending to explain the accelerations of Pioneer 10 and 11 as well, we have to imagine either that $p$ or $a_0$ is variable or that MOND applies to spacecraft but not planets, and then, particularly if $a_0$ is not variable, imagine there will be a way to explain why the data does not show the expected change when the distance of the spacecraft from the sun changes.


\begin{thebibliography}{9}

\bibitem{sanders}  Robert H. Sanders and Stacy S. McGaugh, ``Modified Newtonian dynamics as an alternative to dark matter", Ann. Rev. Astron. Astrophys. {\bf 40}, 263-317 (2002), astro-ph/0204521.

\bibitem{anderson}  J.~D. Anderson, P.~A. Laing, E.~L. Lau. A.~S. Liu, M.~M. Nieto, and S.~G. Turyshev, ``Study of the anomalous acceleration of Pioneer 10 and 11", Phys. Rev. D{\bf 65}, 082004 (2002).
\end{thebibliography}
\end{document}